\documentclass[a4paper,twocolumn]{article}
\renewcommand{\section}[1]{\par\noindent\textbf{#1}\par}
\title{Comment on ``Learning to build the bomb''}
\author{Ahmad Shariati\thanks{Dept of Physics, Alzahra University, Tehran 19935, Iran;
e-mail: \texttt{shariati@mailaps.org}}}
\date{7 Feb 2008}
\begin{document}
\maketitle
According to Alisa Carrigan's opinion in
\textit{Physics Today} Dec 2007 \cite{Carrigan}, to
prevent proliferation of nuclear weapons, certain rules should be set
to prevent the spread of a particular kind of knowledge.  Her argument
goes as follows:
To build nuclear weapons, scientists and engineers of potentially rogue 
countries need to know some technics which could be learnt in nonmilitary peaceful
activities, for example in nuclear power plants.  Therefore, to prevent
some countries access to nuclear weapon knowledge, one should prevent their
scientists and engineers being trained in such facilities.  As
Carrigan says, \textit{knowledge proliferation} is as important as
nuclear proliferation.  To show this, Carrigan mentions the case
of South Africa's nuclear program---some scientists and engineers
having trained in USA and Europe in non-military, peaceful,
academic activities, obtained enough knowledge and expertise to
make their own nuclear weapons.  Carrigan says that the cases of
North Korea, India, and Pakistan obtaining nuclear weapons, and
Iran's progress in uranium enrichment are alike.

I would like to comment on this line of reasoning.

\section{Logical consequences}
First, nuclear
weapons are not the only threats.  Chemical and biological weapons
are as dangerous as nuclear weapons.  So if we accept this logic,
the restriction should not be limited to nuclear physics and the
related fields--- by the same reasons, various fields in
chemistry, chemical engineering, pharmaceutical and biological
sciences, physics, and mechanics must be off-limits.  After that comes
various fields of mathematics, for example number theory, and
software engineering; as they have applications in cryptography.
Just think of a terrorist attack by some hacker to a computer that
is controlling an airlines corridor traffic.  Even quantum
computation is also dangerous, because it has applications in
deciphering. Where should one stop?

Carrigan distinguishes between explicit and tacit knowledge. But
there is no permanent sharp line between explicit and tacit
knowledge. For example, the need to use fabric gloves to
assemble centrifuges, the problem mentioned in Carrigan's article,
now that it is being published, has been transformed from tacit to
explicit.  Since people do have access to explicit knowledge,
through books and journals, it is not sufficient to monitor the
sources of tacit knowledge---to prevent proliferation of the
required knowledge, it is necessary to control the flow of
explicit knowledge as well.  This requires establishing a system
of censorship.

I think the logical consequence of accepting Carrigan's idea is a
kind of ``Knowledge Nonproliferation Treaty''---a system
to monitor and control the flow of information through books,
journals, internet, participation in conferences, sabbaticals,
etc. Such a system, if implemented, simply means this: Humans are
divided into two categories, those having the knowledge of making
nuclear, chemical, and biological weapons, and those that have not
yet this knowledge.  The first category has the right and must do
its best to prevent the second category obtaining the required
knowledge and technology. At this limit, I think, it is nothing
but a variant of apartheid.

An inevitable conclusion in line with Carrigan's arguments would
be that good people should control other people in the sense that
if other people were approaching dangerous knowledge (even by
themselves), good people should prevent them even if necessary by
force, even if necessary by getting rid of the scientists of other
people and destroying their scientific facilities, including their
libraries and equivalent digital resources.  This, simply would
force people in the second category---those who are forbidden to
have the sacred knowledge---to invoke dirty tricks.

\section{Scientific apartheid doesn't work}
I am not saying that scientific apartheid is bad, for valuing something as good or bad could
not be judged scientifically.  What \textit{is} important, I think, is that this scientific
apartheid does not work, and is not a suitable means to establish a sustainable peace.

It does not work, because it is now almost impossible to impose
it. Today, contrary to say 100 years ago, even people in
developing countries do have access to the basics of the
scientific method and the fundamentals of science. Once one knows
these, it is in principle possible to produce the forbidden
knowledge. After all, this is what scientists in the developed
countries have done, and assuming that there is no meaningful
distinction between the intelligence of people in different
countries, if people in say USA have been able to learn or
construct things by themselves, people in other countries can do
that as well, though with some delay.
So a Knowledge Nonproliferation Treaty does not help,
since knowledge is not only transported, but also produced---the
example of fabric gloves mentioned in Carrigan's article is a very
good example of this.

\section{Reducing tensions}
Now let us consider this problem from another point of view. The
case of South Africa's nuclear program is worthy of discussing.
Why South Africa made weapons, and why finally destroyed its
weapons?  I think the answer is that, 4 decades ago South Africa
was a country, having trouble with its neighbors---and its own
people as well.  After the Apartheid era, the troubles being solved,
and now South Africa does not need any nuclear weapons. Which
other countries have made nuclear weapons? North Korea, having
trouble with South Korea; Israel, having trouble with all its
neighbors; Pakistan, having trouble with India; India, trouble
with Pakistan.  What Carrigan points, is that all these nations
were able to obtain the required knowledge, and all of them from
non-military activities.  What I conclude from this, is that if
some nation has enough motivation to build a dangerous weapon, it
probably can obtain the required knowledge---and Carrigan
says that this has always been achieved by native scientists.
\textit{Now, if we want to make a sustainable peace, why not try to reduce
the motivation of nations to have weapons?}

\section{In mathematical terms}
Let me formulate my view more mathematically. Let $K(T)$ be the
probability of nation $X$ to have the knowledge and technology
required to produce a nuclear weapon before time $T$.
Let $H(T)$ be the probability of nation $X$ to have nuclear weapons
before time $T$.
And let $U(T)$ be the probability of nation $X$ using nuclear
weapons before time $T$. For time $T$ let's consider 2020 for the
moment.

One can argue that $K$ is an increasing function of the level of
ease physicists from $X$ can visit foreign universities having
nuclear physics departments. Denote this level of ease with $x$. 
One can also argue that
$H$, and especially $U$ depend critically on the regional
tensions---by region I mean the Middle East, Kashmir, Korean
Peninsula, etc.  Let $y$ denote the level of this tensions.

The most important task is to try to reduce $U$, and after that
$H$. Carrigan is saying that $K$ is an increasing function of $x$,
even though so far all those nations who had enough motivation,
have succeeded in obtaining nuclear weapons.   What I am saying is that we
know that decreasing $y$ has quite profound effects on
reducing $H$ and $U$, and we know that in the only case for which
the regional tensions vanished, the country (South Africa) destroyed
its weapons.  So why not trying to reduce the regional tensions?

Besides, $K(T)$ is obviously an increasing function of time $T$,
because it is an increasing function of the overall level of
knowledge and technology of the world.  Day by day it will become more and more
difficult to make $K$ not approaching 1.
However, for $H(T)$ and especially $U(T)$ it is not obvious that they
are increasing functions of time, for they depend on the
political conditions at times $t \leq T$.  So again, it is
quite wiser to try to reduce the regional tensions.

Finally, trying to reduce the level of knowledge of nation $X$, or
preventing it from increasing its knowledge, by establishing a type
of Knowledge Nonproliferation Treaty,
will cause $X$ to become more aggressive and less
developed.  I think both of these would increase $H(T)$ and
$U(T)$.


\begin{thebibliography}{9}
\bibitem{Carrigan} Alisa L.\ Carrigan, ``Learning to Build the Bomb'', {\it Physics Today}, Dec 2007, pp. 54-55.
\end{thebibliography}
\end{document}